\documentclass[9pt,twocolumn,twoside]{pnas-new}

\templatetype{pnasresearcharticle} 
\usepackage{multirow}
\usepackage{gensymb}
\usepackage{graphicx}
\usepackage[caption=false]{subfig}
\usepackage{times}
\usepackage{color}
\usepackage{soul}
\usepackage[normalem]{ulem}
\usepackage{caption}
\usepackage{array}
\title{Machine learning reveals systematic accumulation of electric current in lead-up to solar flares}

\author[a,1]{Dattaraj B. Dhuri}
\author[a,b]{Shravan M. Hanasoge} 
\author[c,d]{Mark C. M. Cheung}

\affil[a]{Tata Institute of Fundamental Research, Mumbai 400005, India}
\affil[b]{Center for Space Science, New York University Abu Dhabi, Abu Dhabi, United Arab Emirates}
\affil[c]{Lockheed Martin Solar \& Astrophysics Laboratory, Palo Alto, CA 94304, USA}
\affil[d]{Hansen Experimental Physics Laboratory, Stanford University, CA 94305, USA }

\leadauthor{Dhuri} 

\significancestatement{Reliable flare forecasting is essential for improving  preparedness for severe space-weather consequences. Flares also serve as probes of solar magnetic processes and the emergence of flux at the solar surface. Training Machine Learning (ML) algorithms using magnetic-field observations for improving flare forecasting has been extensively studied in prior literature. Instead, here we use ML to understand the underlying mechanisms governing flares. We train ML algorithms to classify flaring and non-flaring ARs with high fidelity and, for the first time, report statistical trends for AR evolution days before and after M- and X-class flares. These trends are interpreted in terms of existing models of sub-surface magnetic field and flux emergence. Our results also provide hypotheses for achieving reliable flare forecasting.}

\authorcontributions{D.B.D performed statistical analysis and contributed to interpretation of results and drafting of the manuscript. S.M.H. and M.C.M.C. designed the study and contributed to interpretation of results and drafting of the manuscript.}
\authordeclaration{The authors declare that they have no competing financial interests.}
\correspondingauthor{\textsuperscript{1}To whom correspondence should be addressed. E-mail: dattaraj.dhuri@tifr.res.in}


\begin{abstract}
Solar flares --- bursts of high-energy radiation responsible for severe space-weather effects --- are a consequence of the occasional destabilization of magnetic fields rooted in active regions (ARs). The complexity of AR evolution is a barrier to a comprehensive understanding of flaring processes and accurate prediction. Though machine learning (ML) has been used to improve flare predictions, the potential for revealing precursors and associated physics has been underexploited. Here, we train ML algorithms to classify between vector-magnetic-field observations from flaring ARs, producing at least one M-/X-class flare, and non-flaring ARs. Analysis of magnetic-field observations accurately classified by the machine presents statistical evidence for (1) ARs persisting in flare-productive states --- characterized by AR area --- for days, before and after M- and X-class flare events, (2) systematic pre-flare build-up of free energy in the form of electric currents, suggesting that associated subsurface magnetic field is twisted, (3) intensification of Maxwell stresses in the corona above newly emerging ARs, days before first flares. These results provide new insights into flare physics and improving flare forecasting.
\end{abstract}

\begin{document}

\maketitle
\ifthenelse{\boolean{shortarticle}}{\ifthenelse{\boolean{singlecolumn}}{\abscontentformatted}{\abscontent}}{}

\dropcap{B}y virtue of buoyancy, magnetic fields generated in the interior of the Sun rise to the photosphere - the visible solar surface -  and emerge as bipolar active regions (ARs) \cite{Cheung2014,Stein2012}. Emerging flux and electric currents energize coronal magnetic field that is rooted in ARs \cite{Leka1996}. Magnetic reconnection occasionally releases free energy built up in the coronal loops in violent events such as solar flares \cite{Shibata2011,Su2013}. M- and X-class flares, producing X-Ray flux $>10^{-5}\ \textrm{W-m}^{-2}$ and $>10^{-4}\ \textrm{W-m}^{-2}$ respectively as measured by {\it Geostationary and Environmental Satellite} (GOES), can have severe space weather consequences \cite{Eastwood2017}. Operational flare forecasts are based on subjective analyses of AR morphology \cite{McIntosh1990,Rust1994,crown2012validation}. Reliable precursors for accurate flare forecasting, however, remain elusive \cite{AllClear}.

The complex nature of AR dynamics hinders straightforward interpretation of flare observations, though AR magnetic-field features related to flare activity are known from case and statistical studies \cite{SCHRIJVER2009739,Leka2008,Wang2015}. Recurrent flares are found to be associated with continuously emerging magnetic flux \cite{Nitta2001}. ARs producing M- and X-class flares contain a prominent high-gradient region separating opposite polarities \cite{Schrijver2007}. Magnetic helicity and electric current is found to be accumulated in ARs prior to major flares \cite{Park2008,Kontogiannis2017}. Minutes before the onset of flares, increased Lorentz forces in ARs are observed as a result of elevated pressure from the coronal magnetic field \cite{Sun2017,Fisher2012}. Such AR features can be quantified using photospheric vector-magnetic-field data \cite{Hoeksema2014} from Helioseismic and Magnetic Imager (HMI \cite{Scherrer2012}) on board Solar Dynamics Observatory (SDO \cite{Pesnell-etall2012}).

ML --- efficient in classifying, recognizing and interpreting patterns in high-dimensional data sets --- have been applied to predict flares using many AR features simultaneously. Such studies are aimed at developing reliable forecasting method and identifying features most relevant to flare activity \cite{Ahmed2013,bobraflareprediction,Florios2018,Jonas2018}, obtaining new AR features that yield better forecasting accuracy \cite{Florios2018,Raboonik2016} and comparing performances of different ML algorithms \cite{Nishizuka2017}. Flare prediction accuracy is expected to depend on forward-looking time i.e.\, how far in advance flares can be predicted. Existing studies, which use AR observations ranging from 1-48 hours prior to flares, however suggest that forecasting accuracy is largely insensitive to forward-looking time \cite{bobraflareprediction,Raboonik2016,Huang2018}. Thus flaring ARs may exist in a flare-productive state long before producing a flare. This motivates the present work where we explicitly train ML algorithms to classify between photospheric magnetic fields of flaring and non-flaring ARs. The trained machine builds a correlation (probability distribution function) between AR photospheric magnetic fields and flaring activity in AR coronal loops. We analyze time evolution of machine correlation between AR magnetic fields and flaring activity to investigate a) whether magnetic fields from flaring and non-flaring ARs are intrinsically different, b) statistical evolution in flaring ARs days before and after flares, as well as c) the development of emerging ARs days before first flares.

\section*{Methods}
We consider ARs between May 2010 - Apr 2016. Using the GOES X-ray flux catalog, we identify ARs that produce at least one M- or X-class flare during its passage across the visible solar disk as flaring and otherwise as non-flaring. We only consider ARs with maximum observed area $>$ 25 Mm$^2$. This restriction serves to eliminate thousands of very small-scale non-flaring ARs and no flaring AR. We represent AR photospheric magnetic fields by 12 features, listed in Table 1a, computed from HMI magnetograms every 12 minutes (SI Appendix, Table S1). These features are publicly available in the data-product Space Weather HMI Active Region Patches (SHARPs) \cite{Bobra2014} and produce optimum flare forecasting performance \cite{bobraflareprediction}. Using ML, we classify whether a given magnetic-field observation, represented by SHARP features, corresponds to a flaring or non-flaring AR. Part of the available data are used to train the machine and validate the performance. The trained machine is then used to classify and analyze magnetic fields of ARs in the remaining unseen data --- the test data. Formally, the trained machine gives an optimum mapping $\textbf{X} \rightarrow Y$ where $\textbf{X}$ is a 12-dimensional SHARP feature vector and $Y \in \{1,0\}$ is the machine prediction. $Y=1$ implies that the AR has flared or is about to, and $Y=0$ implies that the AR belongs to the non-flaring population.  For flaring ARs, SHARP feature vectors that yield $Y=1$ are True Positives (TP) and $Y=0$ are False Negatives (FN). For non-flaring ARs, SHARP feature vectors that yield $Y=0$ are True Negatives (TN) and $Y=1$ are False Positives (FP). We statistically analyze time series of SHARP feature vectors $\textbf{X}(t)$ from TP and FN populations days before and after flares.

\begin{table*}[t]
\centering
\subfloat[]{
\begin{tabular}{lcc}
\toprule
&\textbf{Symbol} & \textbf{Description} \\
\midrule
1 & USFLUX & Total unsigned flux \\
2 & AREA & Area of strong-field pixels in active region \\
3 & TOTUSJH & Total unsigned current helicity \\
4 & TOTPOT & Total photospheric magnetic free-energy density \\
5 & TOTUSJZ & Total unsigned vertical current \\
6 & TOTBSQ & Total magnitude of Lorentz force \\
7 & ABSNJZH & Absolute value of net current helicity \\
8 & SAVNCPP & Sum of modulus of net current of each polarity \\
9 & MEANPOT & Mean photospheric magnetic free energy \\
10 & SHRGT45 & Fraction of area with magnetic field shear $> 45\degree$ \\
11 & R\_VALUE & Sum of flux near polarity inversion line (PIL) \\
12 & TOTFZ & Sum of $z$-component of Lorentz force \\
\bottomrule
\end{tabular}
}\qquad
\subfloat[]{
\begin{tabular}{cccc}
\toprule
 & {\bf Training Data} & \multicolumn{2}{c}{{\bf Test Data}}\\
& I  &  II & III \\
\midrule
\# flaring ARs &85&66&22\\
\# non-flaring ARs &308&273&190\\
\# M-class flares & 304 & 276 & 57 \\
\# X-class flares & 24 & 12 & 1\\
\midrule
\multicolumn{4}{c}{{\bf Nomenclature:}}\\
\multicolumn{4}{c}{{\bf I}: \ May 2010 - Dec 2013 (excluding emerging ARs)}\\
\multicolumn{4}{c}{{\bf II}: \ Jan 2014 - Apr 2016 (excluding emerging ARs)}\\
\multicolumn{4}{c}{{\bf III}: \ May 2010 - Apr 2016 emerging ARs }\\
\bottomrule
\end{tabular}
}
\caption{Data used for classification of flaring and non-flaring active regions (ARs) (a) AR magnetic-field features (SHARPs) used for training ML algorithms. These features are correlated with flare activity, yielding optimum flare-forecasting accuracy \cite{bobraflareprediction,Bobra2014}. (b) Number of ARs and M- and X-class flares considered. Machines are trained using SHARP features from ARs in training and validation data with 10-fold cross-validation. Predictions are made on ARs in the the test data. Emerging ARs are defined as newly appearing within $\pm 60 \degree$ of the central meridian.}
\end{table*}

\section*{Results}
\noindent{{\bf Classification of flaring and non-flaring active regions:}}  We chronologically split the available AR data in two parts. The training and validation data comprises of ARs between May 2010 - Dec 2013 and the test data comprises of ARs between Jan 2014 - Apr 2016. We explicitly study the development of newly emerged ARs, identified from the first recorded observation within $\pm 60\degree$ of the central meridian. Between May 2010 - Apr 2016, only 22 flaring ARs emerged on the visible solar disk,  hence, all `emerging ARs' are included in the test data. The total number of ARs considered are listed in Table 1b. 

 We consider observations from flaring ARs which are within $\pm 72$ hours of M- or X-class flares for training. Note that all magnetic field observations from ARs in the training and validation data are not needed in order to optimally train the machines. Instead, we pick an observation every 96 min from within $\pm 6$ hours of flares and every 864 min otherwise (within $\pm 72$ hours of flares). For non-flaring ARs, we pick an observation every 900 min for training. The choice of these time intervals is inconsequential to the results as long as the number of AR observation samples is adequate for training. For robust training, we apply 10-fold cross-validation. We randomly split the flaring and non-flaring ARs in the training and validation data in 10 parts and use observations from ARs in 9 parts for training and remaining part for validation. This process is repeated 10 times.  Thus, we avoid mixing AR observations in the training and validation sets and thereby avoid artificially boosting the machine performance \cite{Nishizuka2017}. Total number of observations used for training from flaring ARs and non-flaring ARs are 768 and 4323 respectively (SI Appendix, Table S2).
\begin{figure}
    \centering
    \includegraphics[width=\columnwidth]{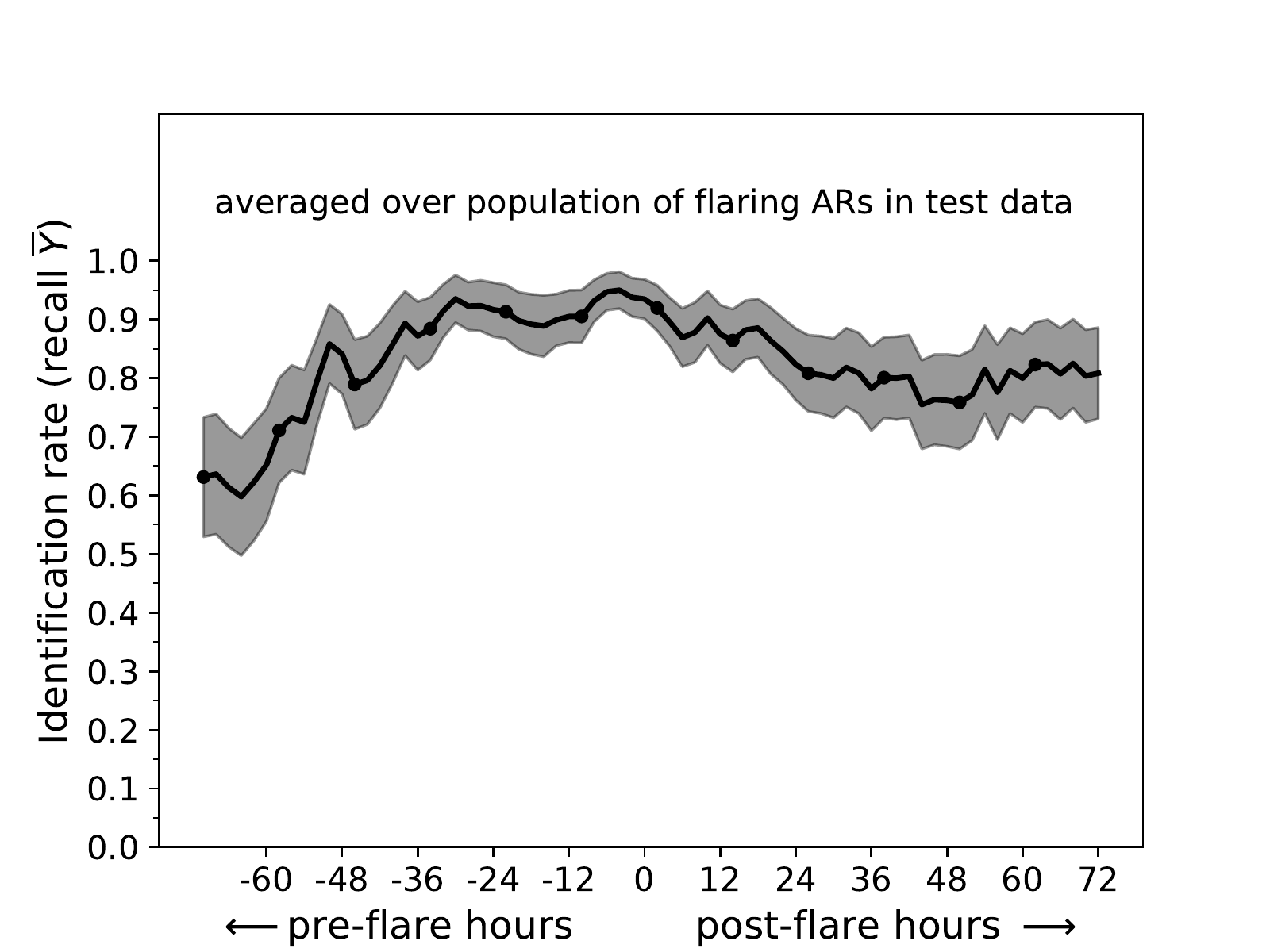}
    \caption{Time evolution of {\it recall}  $\overline{Y}(t_r) = 1/N(t_r) \sum_i^{N(t_r)} Y_i(t_r)$ for flaring ARs in the test data before and after a flare using SVM. The {\it recall} peaks at $\sim 0.9$, $24$ hours prior to flares. For comparison, machine identification error rate ({\it false-positive rate}) for non-flaring ARs in the test data $\langle Y \rangle$ is $\sim 0.1$ (Table 2). Shaded area indicates $1\sigma$ error bar. }
    \label{FIg1}
\end{figure}
\begin{table}[t]
    \centering
   \begin{tabular}{cccc}
\toprule
\multicolumn{4}{c}{{\bf Classification of ARs in the test data using SVM}}\\
\hline
& \multicolumn{2}{c}{{\bf Flaring ARs}} & \multirow{3}{*}{{\bf Non-flaring ARs}} \\
& $> 72$ hours from & $24$ hours before & \\
& flare event & flare event & \\
\hline
\#OBS & 7952 & 34 & 43055 \\
{\it recall}& 0.750 $\pm$ 0.085  & 0.913 $\pm$ 0.046 & 0.889 $\pm$ 0.027 \\ 
\bottomrule
\end{tabular}
\caption{Average prediction for all flaring and non-flaring AR observations in the test data using SVM. {\it Recall} value is high even for flaring AR observations, separated from flare events by more than $72$ hours, and non-flaring AR observations. For flaring AR observations $24$ hours before flares, machine prediction yields {\it recall} of $0.91$ which is comparable with reported results \cite{bobraflareprediction}.}
\label{Table2}
\end{table}

A straightforward performance measure for classification problems is accuracy, defined as the fraction of correctly classified observations i.e. $\textrm{accuracy} = \textrm{(TP + TN)}/\textrm{(TP + FN + FP + TN)}$. However, there are 5 times as many non-flaring as flaring ARs in the training and validation data. Hence, the classification problem considered here is class-imbalanced and accuracy is not useful \cite{bobraflareprediction}. {\it Recall}, defined as the accuracy for each class is a more relevant performance metric. For the positive class i.e. flaring ARs, $\textrm{recall} = \textrm{TP}/\textrm{(TP + FN)}$. Using the training and validation dataset, we compare performance of three ML algorithms --- Logistic Regression (Logit), Support Vector Machines (SVM) and Gradient Boosting (GB) --- for classification of flaring and non-flaring ARs (SI Appendix). SVM yields slightly higher 10-fold cross-validation {\it recall} value $0.83 \pm 0.12$ (SI Appendix). 
\begin{table*}[b]

    \centering
    \begin{tabular}{lcccccc}
    \toprule
    & & \multicolumn{2}{c}{{\bf Flaring ARs}} & \multicolumn{2}{c}{{\bf Non-flaring ARs}} & \\
    & & \multicolumn{2}{c}{(> 72 hours from flare)} & & & \\
    {\bf Symbol} & {\bf Brief Description} & {\bf TP} & {\bf FN} & {\bf FP} & {\bf TN} & {\bf (TP-TN)/$\sigma_{\textrm{TN}}$} \\
    \midrule
    USFLUX ($10^{22}$ Mx) & Total unsigned flux & 3.30 $\pm$ 0.29 & 1.07 $\pm$ 0.11 & 2.48 $\pm$ 0.19 & 0.56 $\pm$ 0.03 & 94.71\\
    TOTUSJH ($10^2$ G$^2$/m) & Total unsigned current helicity & 23.87 $\pm$ 2.09 & 7.89 $\pm$ 0.81 & 19.45 $\pm$ 1.42 & 4.29 $\pm$ 0.21 & 91.27\\
    TOTBSQ ($10^{10}$ G$^2$) & Total Lorentz force & 4.43 $\pm$ 0.41 & 1.53 $\pm$ 0.17 & 3.57 $\pm$ 0.32 & 0.83 $\pm$ 0.04 & 89.10\\
    TOTUSJZ ($10^{13}$ A) & Total unsigned vertical current & 5.43 $\pm$ 0.45 & 1.86 $\pm$ 0.21 & 4.48 $\pm$ 0.35 & 1.00 $\pm$ 0.05 & 85.68\\
    TOTFZ ($10^{23}$ dyne) & Total vertical Lorentz force & -3.30 $\pm$ 0.51 & -0.61 $\pm$ 0.19 & -1.55 $\pm$ 0.20 & -0.34 $\pm$ 0.04 & 78.54\\
    SAVNCPP ($10^{13}$ A) & Sum of net current per polarity & 1.14 $\pm$ 0.11 & 0.39 $\pm$ 0.05 & 0.93 $\pm$ 0.10 & 0.24 $\pm$ 0.01 & 74.70\\
    ABSNJZH (G$^2$/m) & Absolute net current helicity & 254.59 $\pm$ 31.95 & 68.37 $\pm$ 12.66 & 208.28 $\pm$ 28.53 & 40.68 $\pm$ 2.92 & 73.23\\
    TOTPOT ($10^{23}$ erg/cm) & Total magnetic free energy & 5.20 $\pm$ 0.61 & 1.44 $\pm$ 0.30 & 4.80 $\pm$ 0.61 & 0.72 $\pm$ 0.06 & 71.71\\
    AREA (Mm$^2$) & AR area & 262.17 $\pm$ 19.99 & 110.95 $\pm$ 11.88 & 222.82 $\pm$ 18.35 & 62.75 $\pm$ 2.81 & 71.00\\
    R\_VALUE (Mx) & Flux near polarity inversion line & 4.06 $\pm$ 0.09 & 2.81 $\pm$ 0.25 & 4.04 $\pm$ 0.08 & 2.09 $\pm$ 0.09 & 20.95\\
    SHRGT45 (\%) & Area with shear $> 45\degree$ & 29.76 $\pm$ 1.85 & 24.87 $\pm$ 3.81 & 37.30 $\pm$ 2.04 & 20.24 $\pm$ 1.17 & 8.10\\
    MEANPOT ($10^2$ erg/cm$^{3}$) & Mean magnetic free energy & 68.34 $\pm$ 4.61 & 54.82 $\pm$ 10.08 & 81.42 $\pm$ 6.64 & 45.51 $\pm$ 3.04 & 7.51\\
    \bottomrule
    \end{tabular}
    \caption{Average values of SHARP features over flaring and non-flaring AR magnetic field observations categorized by the SVM. True Positives (TP) and  False Negatives (FN) are observations from flaring ARs which are classified as flaring and non-flaring respectively. True Negatives (TN) and False Positives (FP) are observations from non-flaring ARs that are classified as non-flaring and flaring respectively.  }
    \label{tab:my_label}
\end{table*}

\noindent{{\bf Time evolution of machine prediction:}} We are particularly interested in time evolution of magnetic fields in flaring ARs, hence we obtain {\it recall} of the machine prediction on time series of observations from flaring ARs. A time series $\textbf{X}(t)$ of SHARP feature vectors representing continuous AR observations yields a time series of machine prediction $Y(t)$. Flares are known to be temporally clustered \cite{Wheatland2002} and hence we focus on evolution within $\pm72$ hours of flares. We compile time series of observations during a window $t-T_F\in[-72,72] \textrm{ hours}$ centered around a flare event $T_F$. Whenever two consecutive flares on an AR are separated by $< 144$ hours, we split the observations between the flare events in two halves and consider the first half as the post-flare category of the first flare and the second half as belonging to the pre-flare category of the second flare. We align all such time series from flaring ARs at $t-T_F=t_r=0$, the time of flare events, yielding co-temporal $\textbf{X}(t_r)$ and $Y(t_r)$ time series for time $t_r$ with respect to the flare. The machine prediction averaged over the flaring-AR population $\overline{Y}(t_r) = (1/N(t_r)) \sum_{i=1}^{N(t_r)} Y_i(t_r)$ gives instantaneous {\it recall} or identification rate at time $t_r$. Here, $N(t_r)$ is number of magnetic-field observations available at time $t_r$ from the flaring-AR population (SI Appendix, Fig. S2). Thus, {\it recall} $\overline{Y}(t_r)$ is a measure of the time-evolving correlation between SHARP features and flare activity, obtained using the trained machine. Similarly, the machine predictions can be obtained for all observations from non-flaring ARs. Since there is no characteristic time event on non-flaring ARs we find the average machine prediction defined as $\langle Y \rangle = (1/N) \sum_i^N Y_i$. $\langle Y \rangle$ is time and population average over all $N$ non-flaring AR observations and gives {\it false-positive rate}. 

We can now obtain time evolution of machine prediction for flaring ARs in the test data using the trained SVM. Note that none of the observations from the test data were considered during training and cross-validation of the machine, i.e. SVM, performance. Thus all observations in the test data are previously `unseen' by the machine. Similar to the training data, {\it recall} or identification rate is consistently high ($> 0.6$) for days before and after flares for flaring ARs in the test data (Fig. 1). This indicates that flaring ARs persist in a flare-productive state for days before and after flares. With proximity to flares, identification rate increases to a maximum of $0.91$, $24$-hours before flare. This \textit{recall} value is comparable to reported results of flare forecasting using ML \cite{bobraflareprediction} and significantly higher than {\it recall} $\sim 0.55$ obtained through operational forecasts based on subjective AR analyses (as estimated by \cite{crown2012validation}). 
\begin{figure}[t]
    \centering
  \includegraphics[width=0.3\textwidth,trim={2cm 0 2cm 2cm},clip]{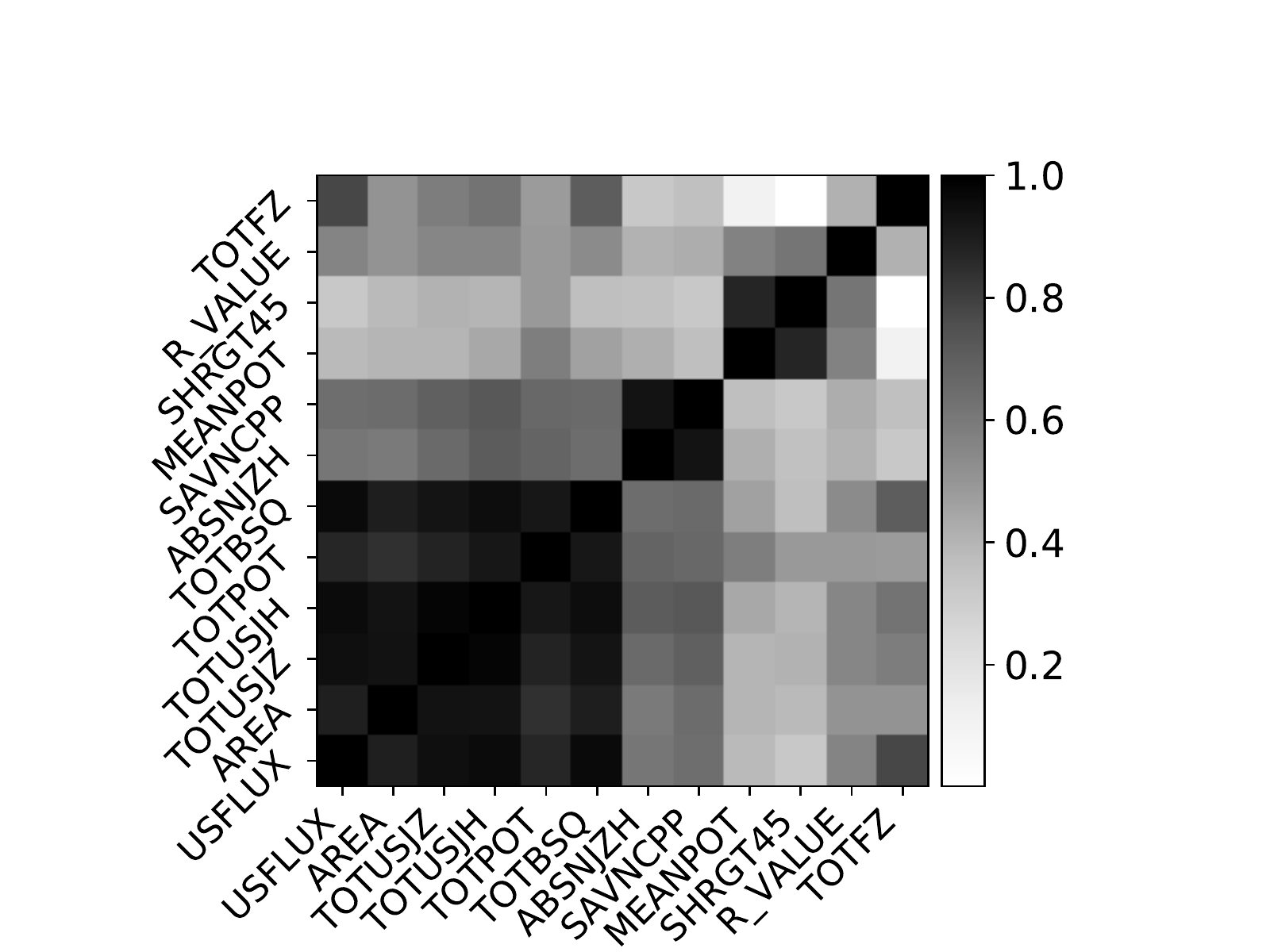}
    \caption{Pearson correlation matrix for SHARP features (see Table 1a for description). Based on the degree of correlation, SHARP features group together in categories representing i) AR magnetic field scale ii) AR energy build-up iii) AR non-potentiality iv) Schrijver R value and v) Lorentz force on AR. P-value of correlation between Total vertical Lorentz force (TOTFZ) and R value is 0.09. All other p-values are $\ll 0.001$.}
    \label{FIg2}
\end{figure}

\begin{figure*}[b]
\centering
\textbf{averaged over population of flaring ARs}\\
\subfloat{\includegraphics[width= 0.33\textwidth,trim={0 1.3cm 0 0},clip]{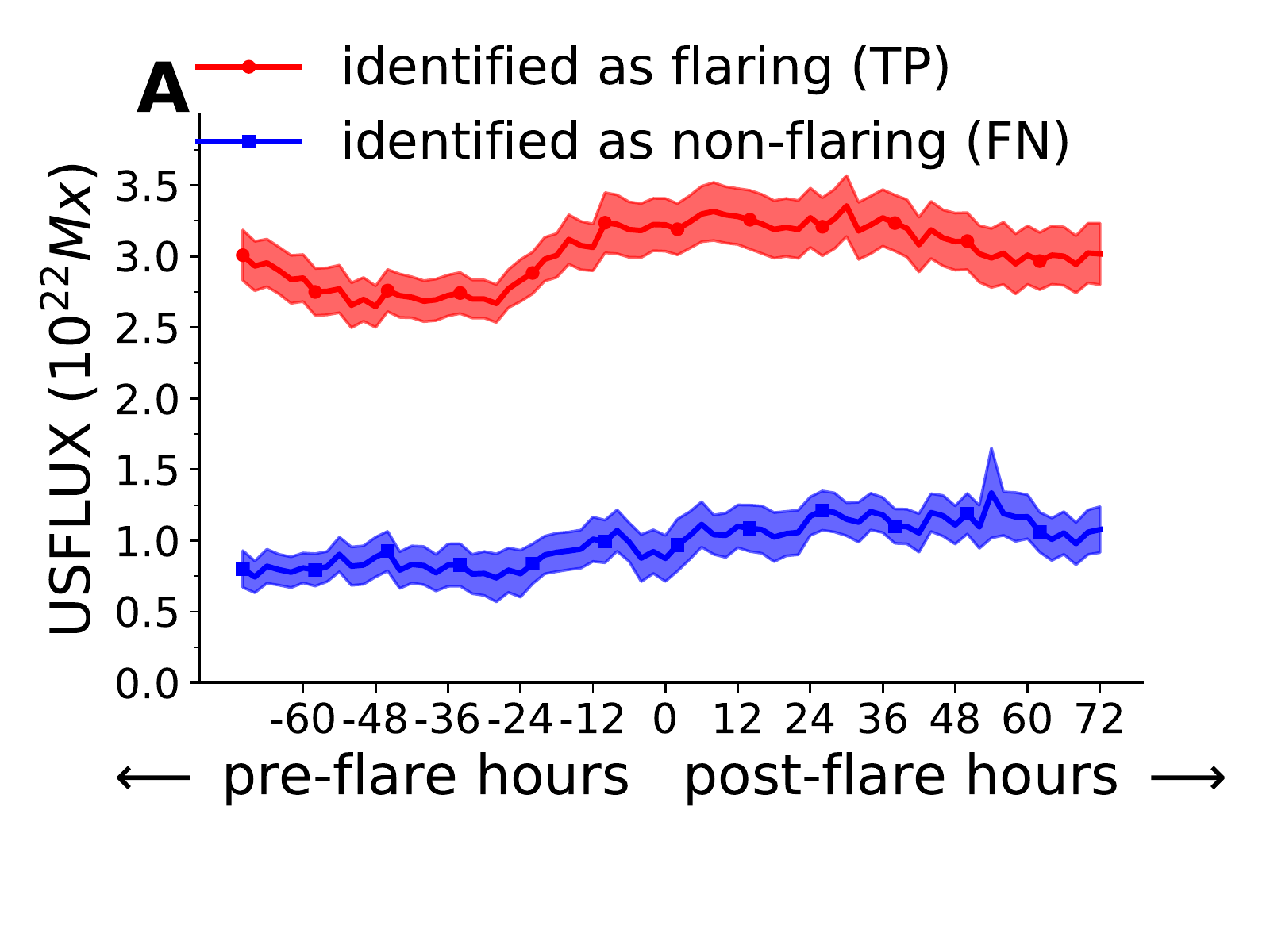}}
\subfloat{\includegraphics[width= 0.33\textwidth,trim={0 1.3cm 0 0},clip]{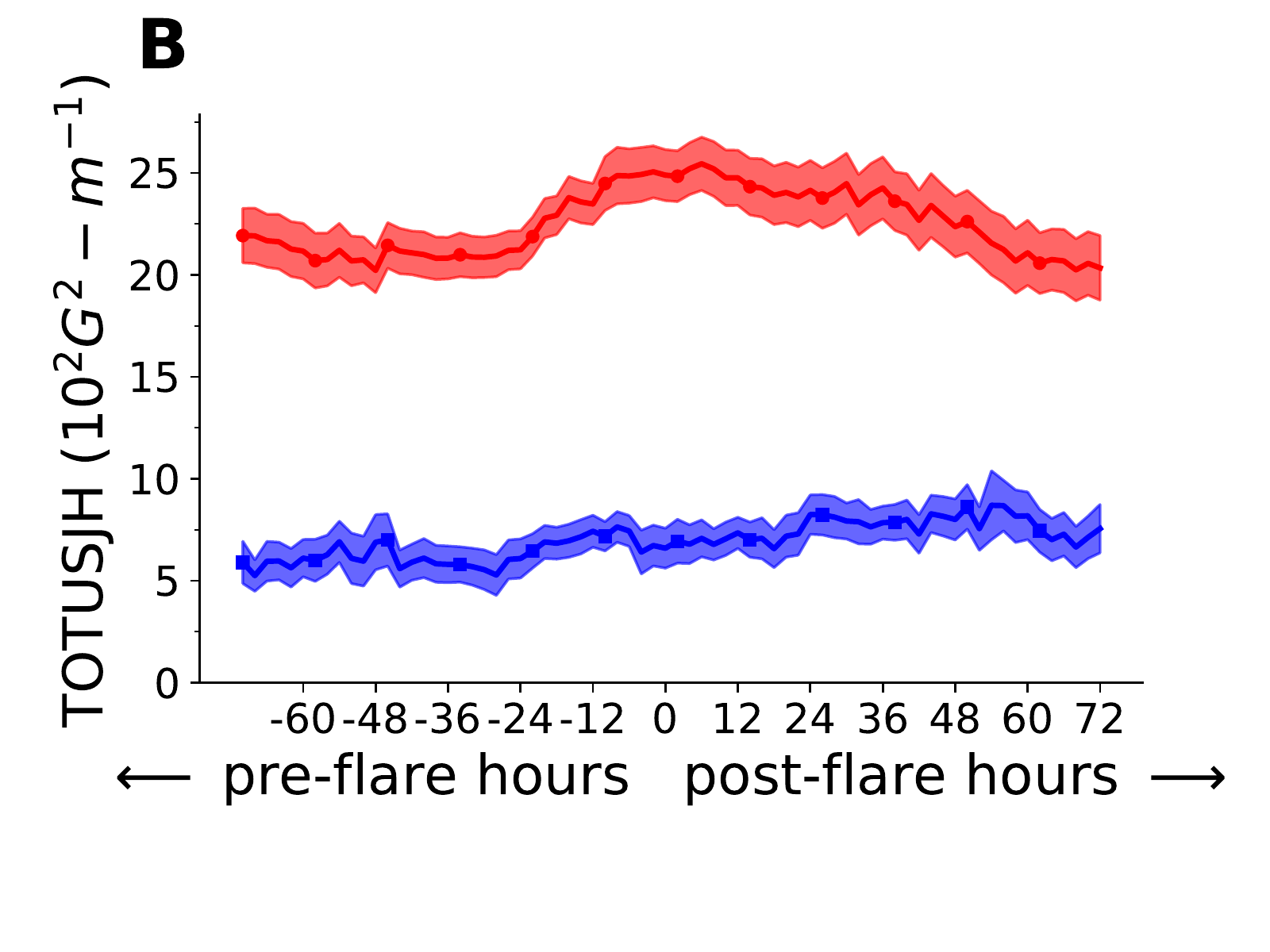}}
\subfloat{\includegraphics[width= 0.33\textwidth,trim={0 1.3cm 0 0},clip]{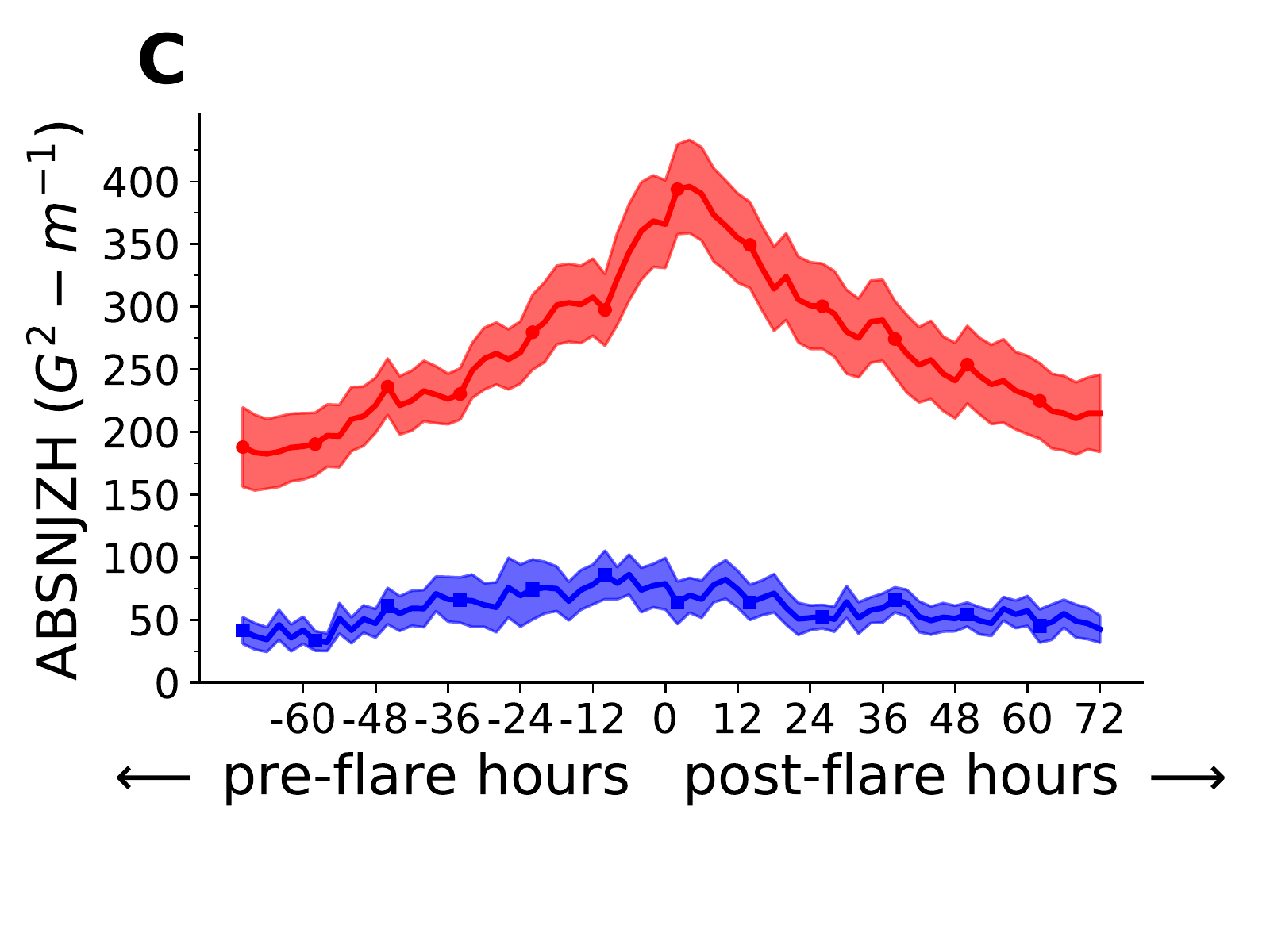}}\\
\subfloat{\includegraphics[width= 0.33\textwidth,trim={0 1.3cm 0 0},clip]{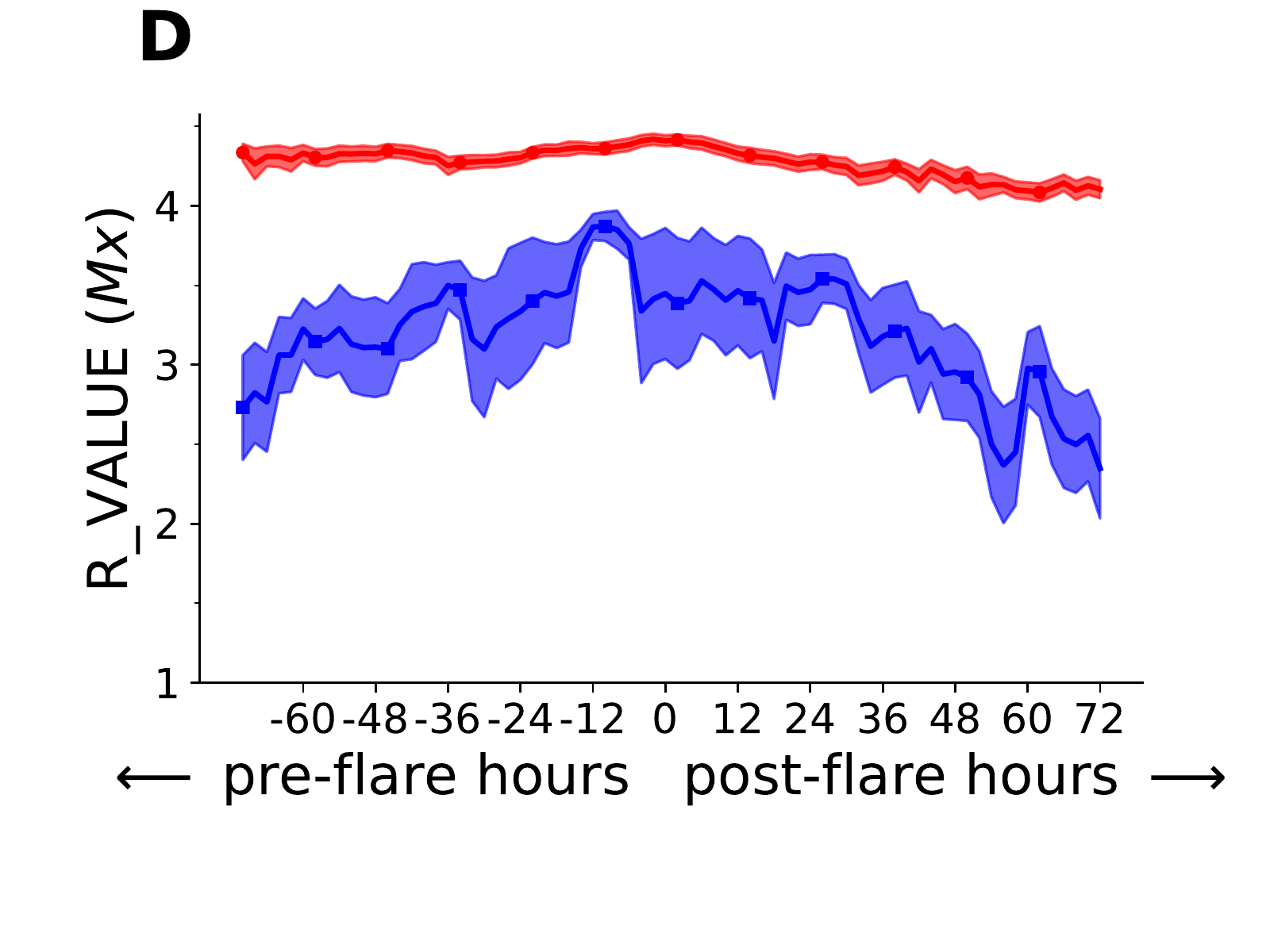}}
\subfloat{\includegraphics[width= 0.33\textwidth,trim={0 1.3cm 0 0},clip]{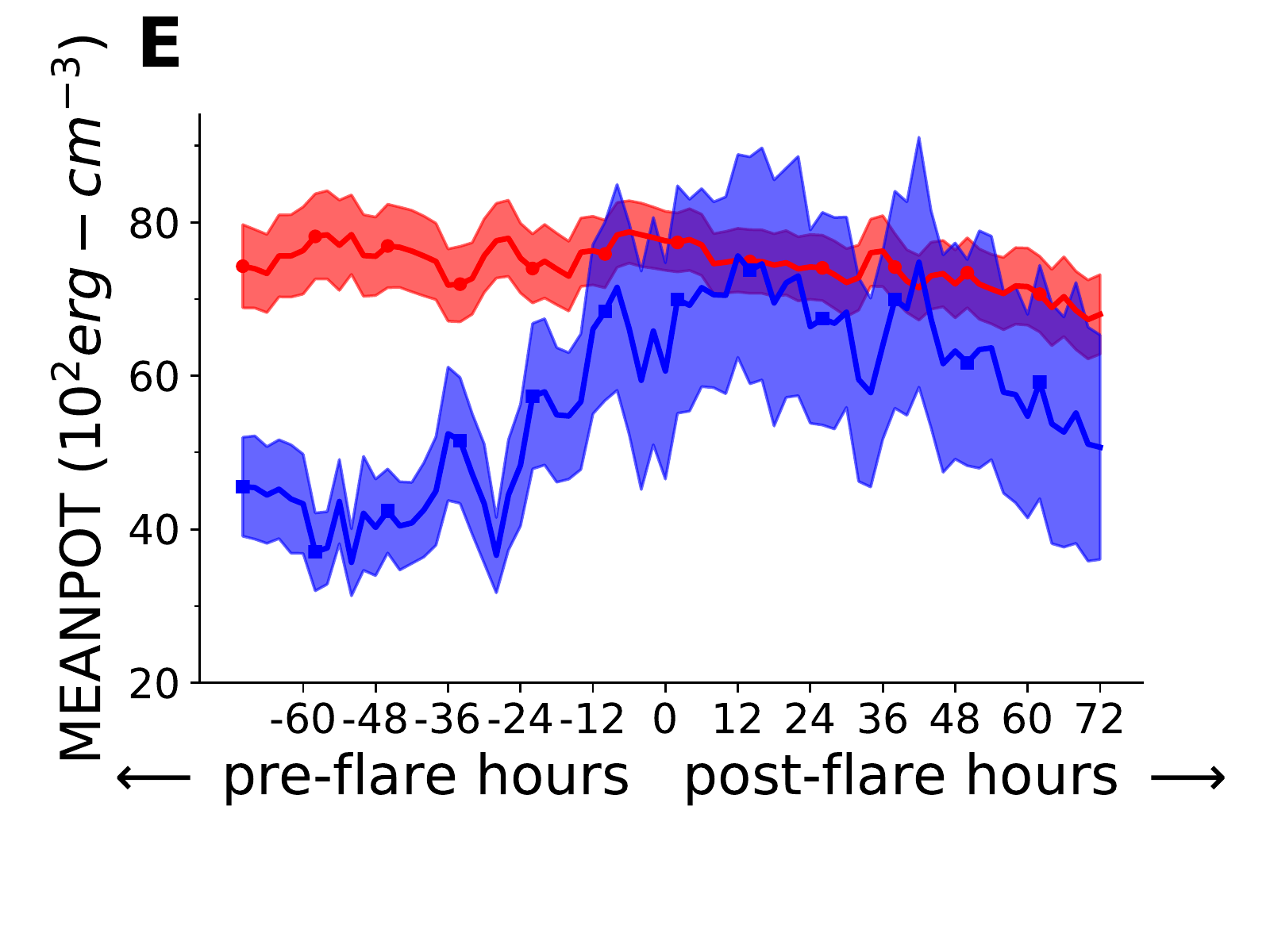}}
\subfloat{\includegraphics[width= 0.33\textwidth,trim={0 1.3cm 0 0},clip]{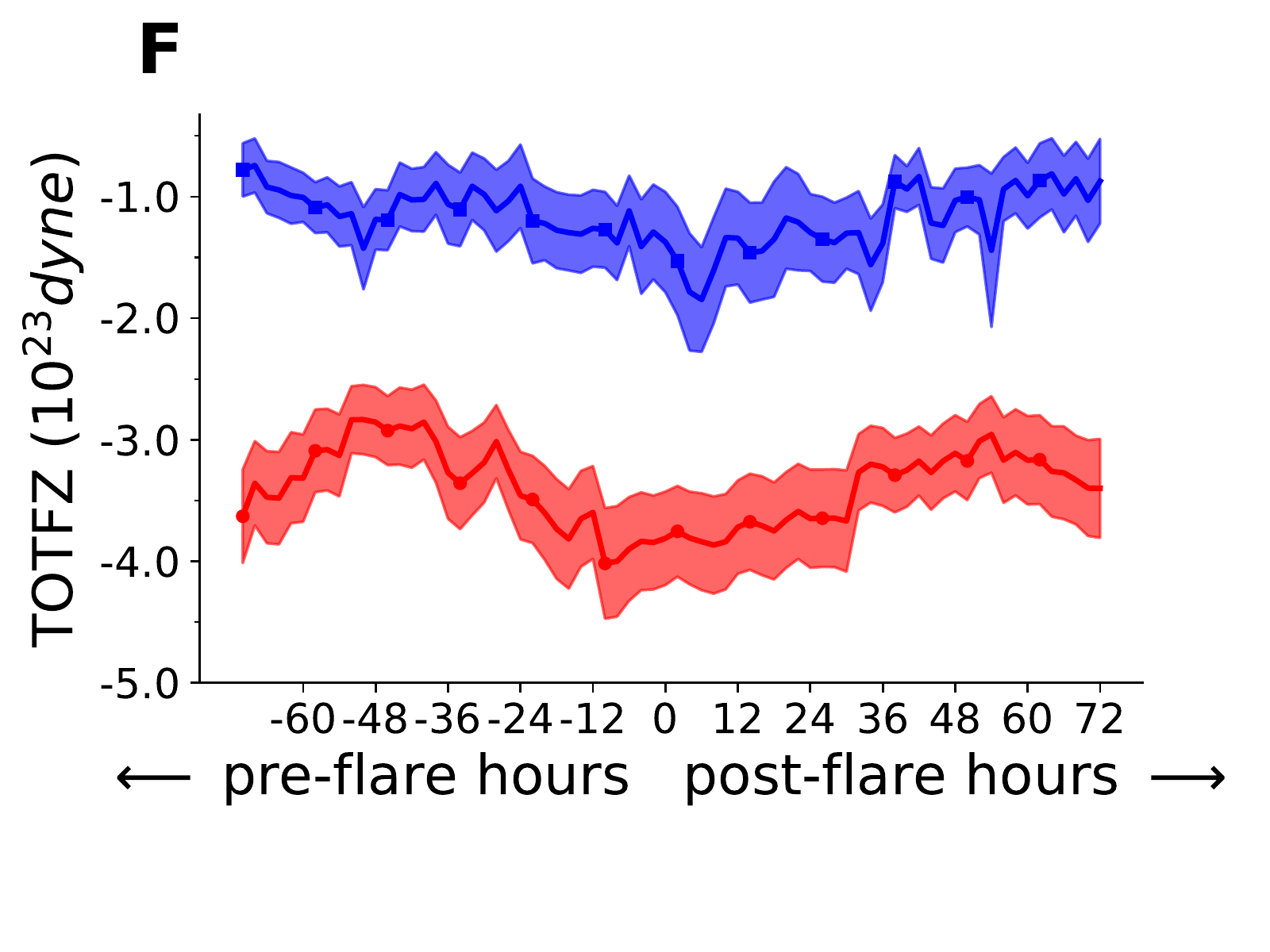}}
    \caption{Time evolution of population-averaged values of SHARP features (see Table 1a for description) before and after flares. Average SHARP feature values over True Positive (TP) and False Negative (FN) flaring AR observations within $\pm 72$-hours of flare events are obtained. Average TP value of total unsigned flux (A) and total unsigned current helicity (B) remains approximately constant before and after flares. Average TP value of absolute net current helicity (C) shows a characteristic steady pre-flare increase and decrease post flare. Sum of flux near polarity inversion line (R-value) (D) also shows a distinct average TP value. For AR non-potential energy (E), average FN value undergoes a sharp increase $\sim 12$ hours before flares. Average TP value of vertical Lorentz force (F) shows continuous pre-flare decrease i.e. increase in downward-directed Lorentz force on AR. Shaded area indicates $1\sigma$ error bars. Legend in (A) applies to sub-figures (A-F).}
\end{figure*}

The number of observations separated from flares by $>72$ hours, reduce significantly to continue time evolution analysis beyond $72$ hours before and after flares (SI Appendix, Fig. S2). Hence, we obtain time- and population-averaged machine prediction ${\langle Y \rangle}_{flaring} = (1/N_{flaring}) \sum_i^{N_{flaring}} Y_i$, where $N_{flaring}$ is number of all flaring AR observations separated from flares $>72$ hours.  ${\langle Y \rangle}_{flaring}$ thus gives {\it recall} for such flaring AR observations. Similarly, {\it recall} for non-flaring ARs is obtained by $1-\langle Y \rangle_{non-flaring}$. Time- and population-average values of {\it recall} for flaring and non-flaring ARs are reported in Table 2. The high value of {\it recall} $\sim 0.75$ even for observations separated from flares $>72$ hours suggests that SHARP features derived from magnetic fields of flaring ARs are statistically significantly different from non-flaring ARs.

\noindent{{\bf Evolution of magnetic fields in flaring active regions}} We have trained an SVM to distinguish between SHARP features derived from magnetic fields in flaring and non-flaring ARs with high fidelity. To understand magnetic field evolution in ARs, we analyze TP and FN populations from flaring ARs and TN and FP populations from non-flaring ARs, as categorized by the machine. We include SHARP features from all ARs in the training and validation data as well as the test data. In Table 3, time- and population-average values of SHARP features over flaring AR observations separated from flares $>72$ hours and non-flaring AR observations are listed. As expected, average TP (also FP) values are strikingly higher than average TN (also FN) values. This difference is listed in terms of standard deviation of average TN values for each of the SHARP parameters, in the last column in Table 3. Total unsigned flux (USFLUX) and total unsigned current helicity (TOTUSJH) are leading contributors to machine classification. Whereas, mean free energy (MEANPOT) and area with shear $>45\degree$ (SHRGT45) minimally influence the classification. Also, SHARP features that lead classification between flaring and non-flaring ARs are an extensive measure of AR magnetic field.

Categories of SHARP features are further highlighted by the Pearson correlation matrix in Fig. 2. Strongly correlated features are divided in the following groups a) extensive features: area, total unsigned flux, total free energy, total Lorentz force, total unsigned vertical current and total unsigned current helicity, b) features that scale with electric current in AR: absolute net current helicity and sum of net current per polarity, c) measures of AR non-potential energy: mean free energy and area with shear $>45\degree$ d) sum of flux near polarity inversion line \cite{Schrijver2007} and e) vertical Lorentz force on AR. From Table 3, we see that the extensive features dominate machine classification, followed by the features that scale with electric current. Meanwhile, features that scale with AR mean non-potential energy contribute the least.

 SHARP features from each of the groups above characteristically evolve before and after flares. For the $m$-th entry of each SHARP feature vector, we calculate the time evolution of population-averaged value $\overline{X^m}(t_r)$, before and after flares, over TP and FN flaring AR observations. SHARP features that scale with AR size are significantly correlated with flare activity. However, similar to total unsigned magnetic flux (Fig. 3A) and total unsigned current helicity (Fig. 3B), average TP values of these SHARP features remain approximately constant before and after flaring and thus characterize flaring AR populations. The average TP value of absolute net current helicity (and also sum of net current per polarity) systematically increases by about two times during the lead up to the flare and decreases subsequently (Fig. 3C). This implies that free-energy build-up in large-scale ARs, manifested in field measurements in the form of photospheric electric current, is dominantly responsible for flares \cite{Shibata2011,Kontogiannis2017}. The high, distinct average-TP value of flux in the neighborhood of the magnetic polarity inversion line (Fig. 3D) is also a striking feature of flaring ARs \cite{Schrijver2007}. AR-associated non-potential energy, which is weakly correlated with electric current (Fig. 2), is not a leading criterion to discriminate between flaring and non-flaring ARs (Fig. 3E).  However, average FN value of non-potential energy shows a sharp increase hours before flare. Average TP value of Total vertical Lorentz force (Fig. 3F) also systematically decreases from days before flares.
\begin{figure*}[t]
\centering
\textbf{averaged over population of emerging flaring ARs}\\
\subfloat{\includegraphics[width= 0.33\textwidth,trim={0 1.3cm 0 0},clip]{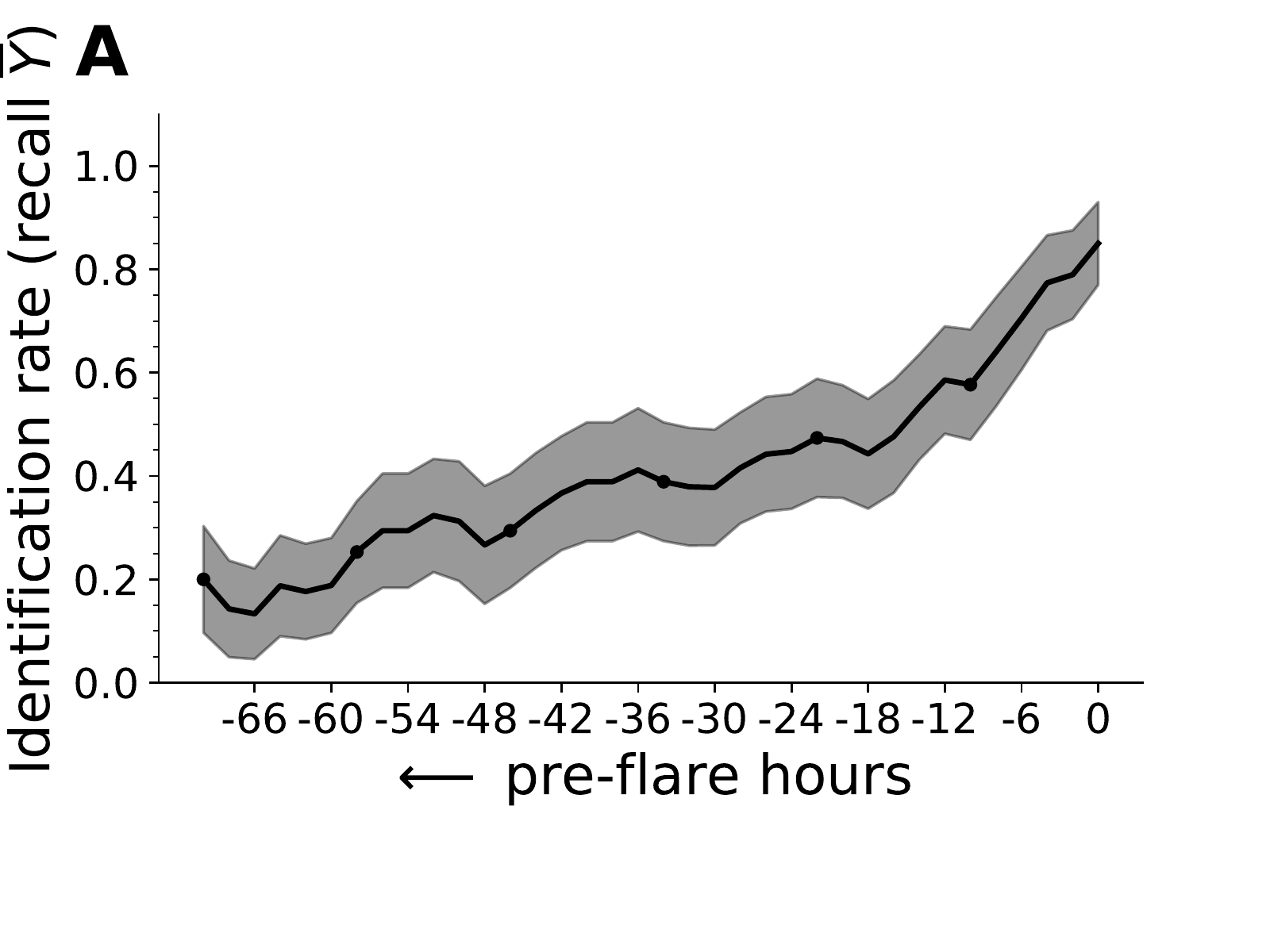}}
\subfloat{\includegraphics[width= 0.33\textwidth,trim={0 1.3cm 0 0},clip]{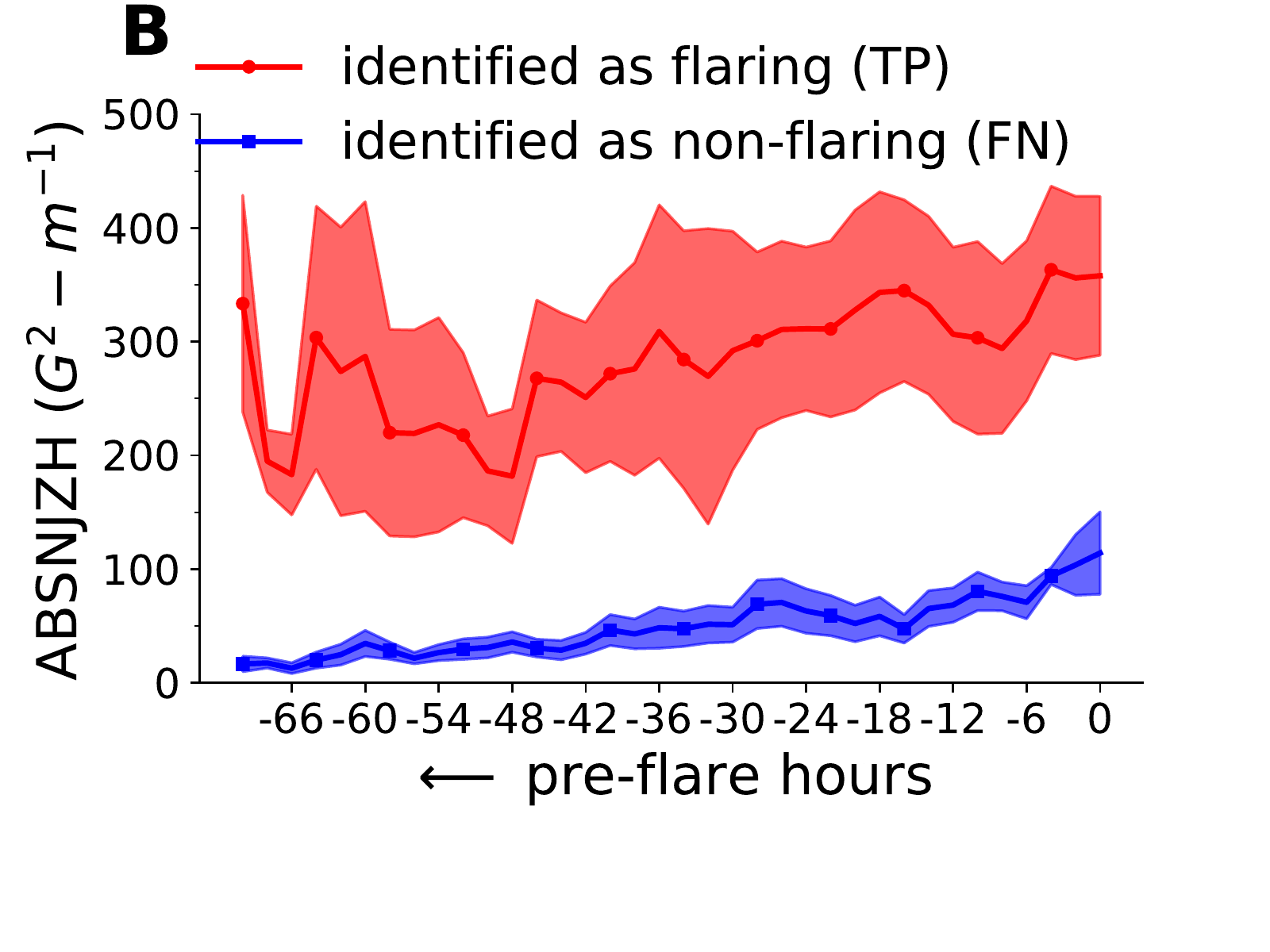}}
\subfloat{\includegraphics[width= 0.33\textwidth,trim={0 1.3cm 0 0},clip]{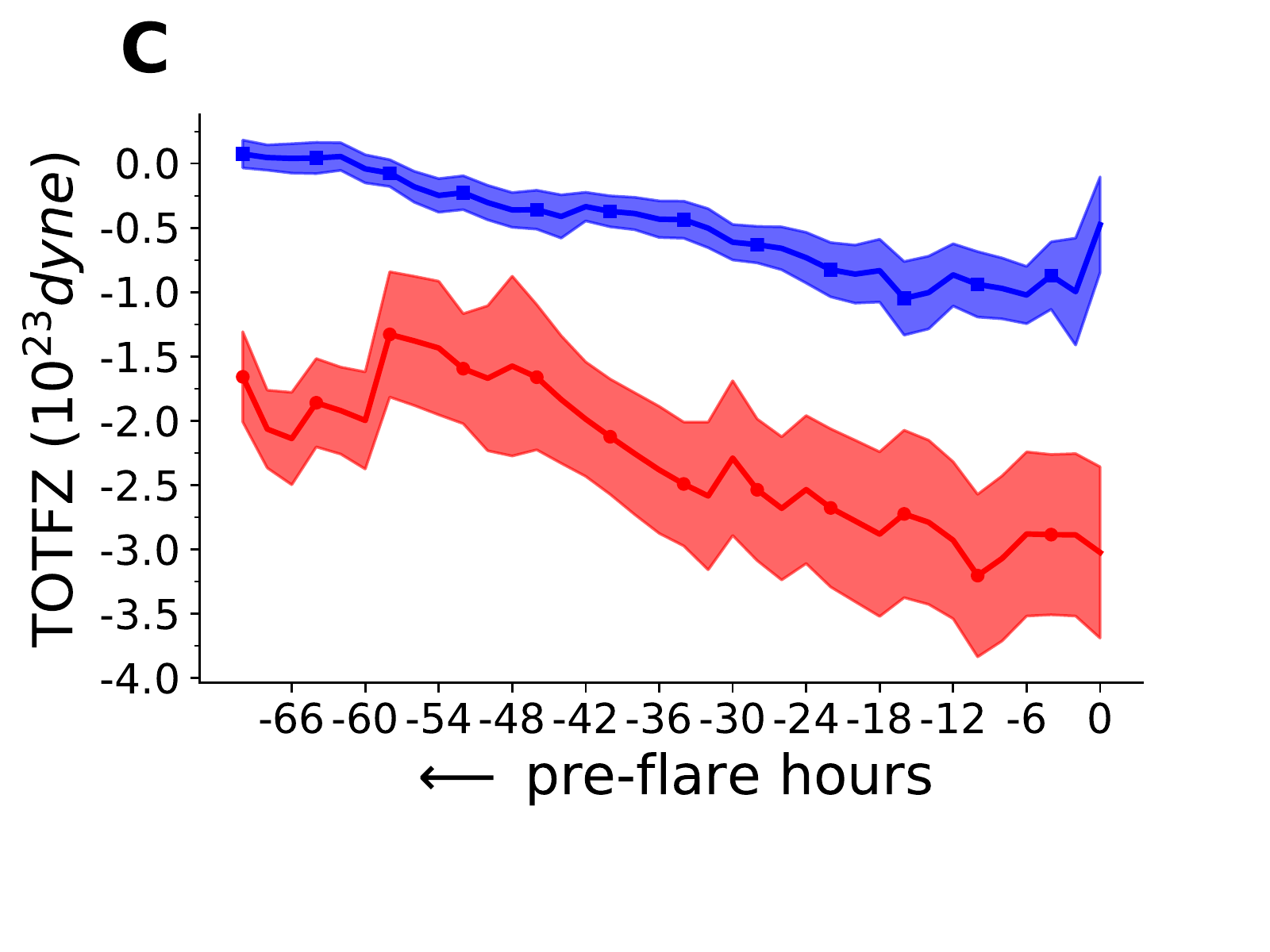}}\\
\subfloat{\includegraphics[width= 0.33\textwidth,trim={0 1.3cm 0 0},clip]{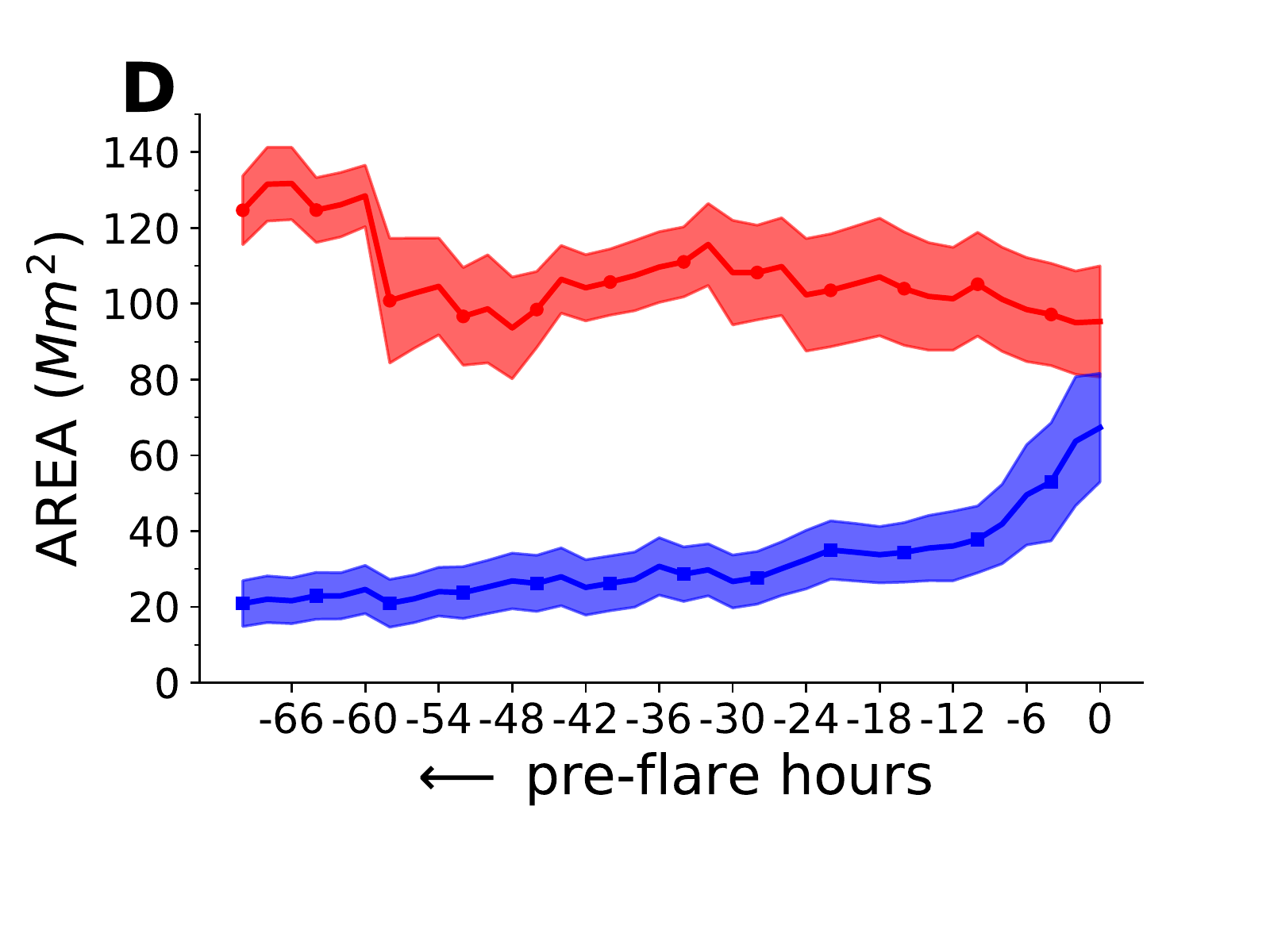}}
\subfloat{\includegraphics[width= 0.33\textwidth,trim={0 1.3cm 0 0},clip]{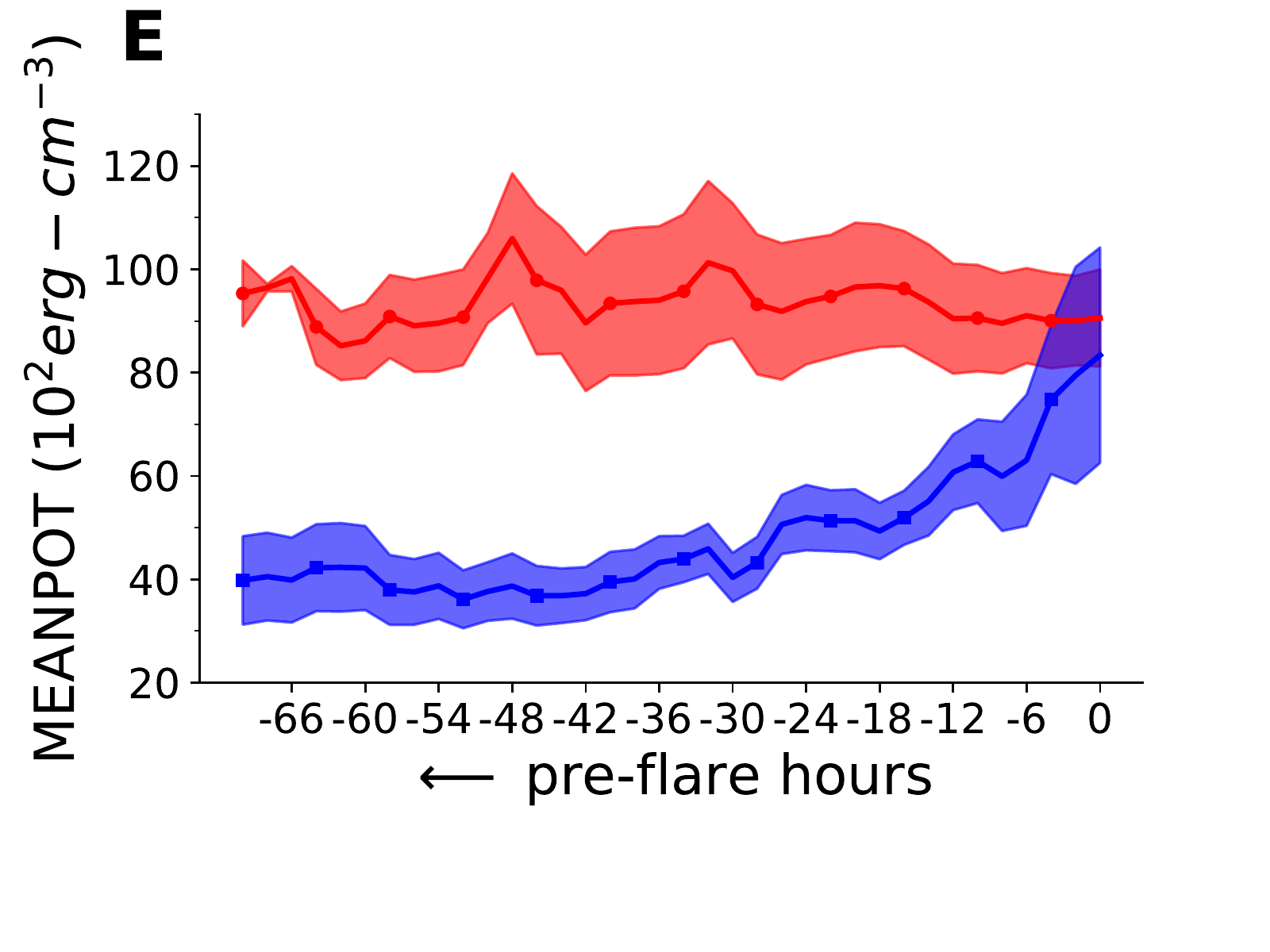}}
\subfloat{\includegraphics[width= 0.33\textwidth,trim={0 1.3cm 0 0},clip]{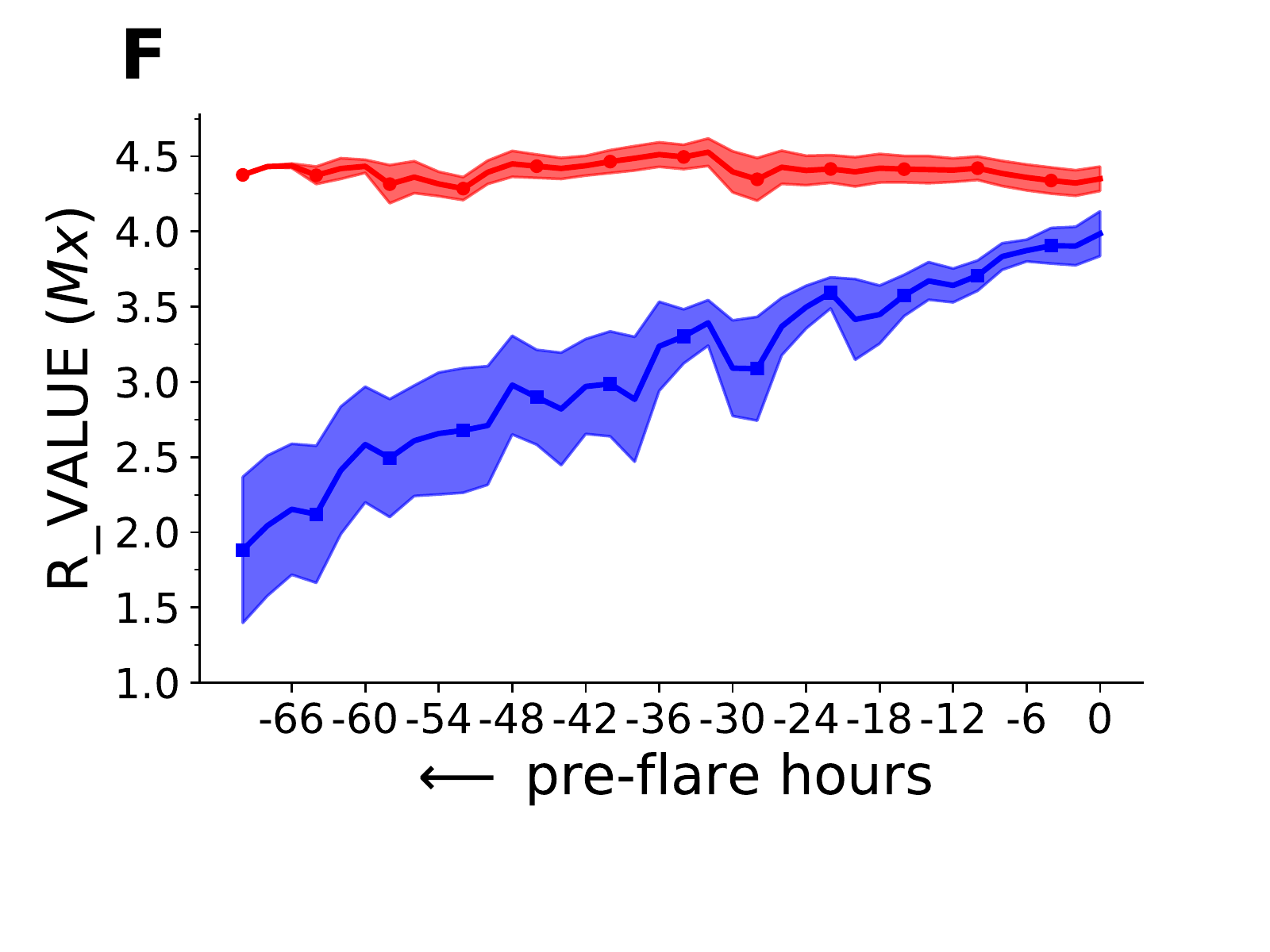}}
    \caption{Machine identification and time evolution of SHARP features for emerging flaring ARs. These emerging ARs are first observed within $\pm 60 \degree$ of the central meridian. (A) Evolution of population averaged identification rate or {\it recall} $\overline{Y}(t_r) = 1/N(t_r) \sum_{i}^{N(t_r)} Y_i(t_r)$ over emerging flaring ARs before $72$-hours of first flare is shown. For comparison, machine identification error rate ({\it false-positive rate}) $\langle Y \rangle$ for emerging non-flaring ARs is $\sim 0.1$. (B-F) Time evolution of population averaged values of SHARP features (see Table 1a for description) over True Positive  (TP) and False Negative (FN) emerging flaring AR observations $72$ hours before first flare. (B) Average TP value of absolute net current helicity increases steadily before flares. (C) Average TP value of vertical Lorentz force on ARs systematically decreases before flares i.e. increase in downward-directed Lorentz force. Average FN values of area (D) and non-potential energy (E) increase sharply hours before flares. (F) Average FN value of R-value increases gradually before flares. Shaded area indicates $1\sigma$ error bars. Legend in (B) applies to sub-figures (B-F).}
\end{figure*}

\noindent{{\bf Development of emerging flaring ARs}} Our analysis shows that extensive SHARP features characteristically distinguish flaring and non-flaring AR populations. Also, values of the extensive features remain approximately constant days before and after flares. On the contrary, newly emerged ARs must start with small values of the extensive SHARP features. Therefore, we are interested in understanding how emerging ARs transition to flare-productive states prior to the first flare. For emerging flaring ARs, we compile observations in a time span $t_r \in [-72,0]$ hours where $t_r=t-T_F$ is time with respect to the first-flare event $T_F$ and compute {\it recall} $\overline{Y}(t_r)$ and TP and FN population-average values for each $m$-th SHARP feature $\overline{X^m}(t_r)$. We see that the machine identification or {\it recall} of newly emerged ARs steadily improves with time and yields maximum {\it recall} value of $\sim 0.7$, $6$ hours before first flares (Fig. 4A). In comparison, the {\it false-positive rate} for emerging non-flaring ARs is $\sim 0.1$ ({\it recall} $\sim$ 0.9). For emerging flaring ARs, population-averaged TP value of absolute net-current helicity (Fig. 4B) shows steady increase, albeit the errorbars are significant. Most notably, the population-averaged TP value of vertically downward-directed Lorentz force increases continuously from days before the flare (Fig. 4C). This may be interpreted as evidence of Maxwell-stress build up in the corona above flaring active regions, which imparts an enhanced downward-directed Lorentz force on the photosphere. For FNs in the emerging flaring AR observations, area (Fig. 4D) and mean non-potential energy (Fig. 4E) show marked increase hours before flare. 

\section*{Discussion} 
We have trained an SVM to classify SHARP features derived from magnetic fields of flaring and non-flaring ARs. The SHARP features used for training (Table 1A) include extensive AR magnetic field features, features that scale with electric current in ARs representing energy build-up, features that scale with AR non-potential energy, flux near polarity inversion line and vertical Lorentz force on ARs. The trained machine classifies flaring AR observations, separated from flare events $>72$ hours, with an average {\it recall} of $0.75$ and non-flaring AR observations with an average {\it recall} of $0.89$. We compare time and population averaged values of TP (FN) observations from flaring ARs, separated from flares $>72$ hours, and TN (FP) observations from non-flaring ARs. We find that extensive AR quantities are leading contributors to the machine classification followed by AR features that scale with electric current. Features derived from AR non-potential energy contribute the least.

A time series of AR magnetic field observations in the form of SHARP features $\textbf{X}(t)$ when fed into the trained SVM results in machine prediction $Y(t)$. Average machine prediction at instant $t_r$, with respect to flares, over population of flaring ARs gives instantaneous {\it recall} $\overline{Y}(t_r)$. $\overline{Y}(t_r)$ can be interpreted as time evolving correlation of SHARP features with flare activity. We find that instantaneous {\it recall} is consistently high, $>0.6$, for flaring AR observations from $72$ hours prior, increasing to a maximum of $0.91$, $24$ hours before the flare. The {\it recall} remains high post flare, suggesting that the ARs lie in a flare-productive state days before and after flares. 

Since the machine prediction $\overline{Y}(t_r)$ is a measure of correlation between SHARP features and flare activity, the temporal evolution of features from accurately classified flaring AR observations, i.e. the TP population, reveals precursors to M- and X-class flares. Similarly, the statistical evolution of inaccurately classified flaring AR observations, i.e. the FN population, has trends that the machine fails to capture. We find that average TP value of extensive AR features --- such as area, total unsigned magnetic flux, total unsigned current helicity --- and flux near the polarity inversion line remain constant for days before and after flares, characterizing flare-productive states for ARs. Total unsigned current helicity is reported to be one of the most significant factors for flare forecasting using machine learning \cite{bobraflareprediction} and is a leading contributor for the classification of flaring and non-flaring ARs as well. However, we find that the key signature of an imminent flare is the systematic build-up of electric currents over days as measured by absolute net current helicity and the sum of net current per polarity. This storage and release of electric current at the photosphere suggests that the sub-surface field associated with flaring ARs is twisted \cite{Cheung2014,Longcope2000}. From case-studies of individual ARs, electric current is known to accumulate prior to major flares \cite{Park2008,Kontogiannis2017}. However, to our best knowledge, this is the first time such clear trends have been observed for days before and after flares, and over statistics of large numbers of ARs.  We show that newly emerging ARs gradually transition to flare-productive states prior to their first flares.  The Lorentz force was hitherto known to increase significantly only minutes before flares \cite{Sun2017}. We find, most notably in the emerging ARs before the first flares, evidence of elevated Lorentz forces exerted on the photosphere by magnetic field in the overlying corona for days before flares.  

This work demonstrates the importance of testing the machine on samples from ARs that are not part of training. Such a restriction is not explicitly imposed in any prior work related to flare forecasting using ML (e.g. \cite{bobraflareprediction,Nishizuka2017,Jonas2018}). Here, we show that SHARP features corresponding to extensive AR quantities (such as total unsigned flux, area etc.) are leading contributors to the machine classification and that the average values of these SHARP features do not change appreciably over a timescale of a few days. Machines trained on observations from a set of ARs, and then tested on observations from the same ARs (albeit for different flares), is likely to have higher {\it recall}  because it has already added to its memory the information it saw in training, namely a similar set of SHARP features. Hence, for accurate testing of the machine, it is important that training and test data do not contain observations from the same ARs.

Class-imbalance between flaring and non-flaring ARs implies that even a false-positive rate of $\sim 0.1$ leads to a significant number of non-flaring ARs being classified as flaring. These FP magnetic fields are from large-scale ($\sim 200 \textrm{Mm}^2$) non-flaring ARs with high values of extensive AR features.  Moreover, FN magnetic fields are from small-scale ARs ($\sim 100 \textrm{Mm}^2)$. We see that average FN value of non-potential energy shows a sharp increase hours prior to the flare (Fig. 3E and Fig. 4E), possibly caused by rapidly emerging flux (Fig. 4D). These pre-flare temporal patterns in small-scale flaring ARs may be accurately captured by ML algorithms trained explicitly on time series data \cite{hamdi}. Thus, achieving reliable flare forecasting requires looking beyond extensive AR features, and focusing on signatures of electric current build-up and rapidly emerging flux. 

\acknow{D.B.D is thankful to Andr\'es Mu\~noz-Jaramillo and Monica Bobra for insightful discussions.  S.M.H acknowledges funding from the Ramanujan fellowship, the Max-Planck partner group program and the Center for Space Science, New York University, Abu Dhabi. Computing was performed on the SEISMO cluster at the Tata Institute of Fundamental Research.}

\showacknow{}

%\bibliography{pnas-sample}

\begin{thebibliography}{10}

\bibitem{Cheung2014}
Cheung MCM, Isobe H (2014) Flux emergence (theory).
\newblock {\em Living Reviews in Solar Physics} 11(1):3.

\bibitem{Stein2012}
Stein RF (2012) Solar surface magneto-convection.
\newblock {\em Living Reviews in Solar Physics} 9(1):4.

\bibitem{Leka1996}
Leka KD, Canfield RC, McClymont AN, van Driel-Gesztelyi L (1996) Evidence for
  current-carrying emerging flux.
\newblock {\em The Astrophysical Journal} 462(1):547.

\bibitem{Shibata2011}
Shibata K, Magara T (2011) Solar flares: Magnetohydrodynamic processes.
\newblock {\em Living Reviews in Solar Physics} 8(1):6.

\bibitem{Su2013}
Su Y, et~al. (2013) Imaging coronal magnetic-field reconnection in a solar flare.
\newblock {\em Nature Physics} 9:489.

\bibitem{Eastwood2017}
Eastwood JP, et~al. (2017) The economic impact of space weather: Where do we stand?
\newblock {\em Risk Analysis} 37(2):206--218.

\bibitem{McIntosh1990}
McIntosh PS (1990) The classification of sunspot groups.
\newblock {\em Solar Physics} 125(2):251--267.

\bibitem{Rust1994}
Rust DM, et~al. (1994) Preflare state.
\newblock {\em Solar Physics} 153(1):1--17.

\bibitem{crown2012validation}
Crown MD (2012) Validation of the NOAA space weather prediction center's solar flare forecasting look‐up table and forecaster‐issued probabilities.
\newblock {\em Space Weather} 10(6).

\bibitem{AllClear}
Barnes G, et~al. (2016) A comparison of flare forecasting methods. i. results from the “all-clear” workshop.
\newblock {\em The Astrophysical Journal} 829(2):89.

\bibitem{SCHRIJVER2009739}
Schrijver CJ (2009) Driving major solar flares and eruptions: A review.
\newblock {\em Advances in Space Research} 43(5):739 -- 755.

\bibitem{Leka2008}
Leka KD, Barnes G (2007) Photospheric magnetic field properties of flaring  versus flare-quiet active regions. iv. a statistically significant sample.
\newblock {\em The Astrophysical Journal} 656(2):1173.

\bibitem{Wang2015}
Wang H, Liu C (2015) Structure and evolution of magnetic fields associated with solar eruptions.
\newblock {\em Research in Astronomy and Astrophysics} 15(2):145.

\bibitem{Nitta2001}
Nitta NV, Hudson HS (2001) Recurrent flare/cme events from an emerging flux region.
\newblock {\em Geophysical Research Letters} 28(19):3801--3804.

\bibitem{Schrijver2007}
Schrijver CJ (2007) A characteristic magnetic field pattern associated with all
  major solar flares and its use in flare forecasting.
\newblock {\em The Astrophysical Journal Letters} 655(2):L117.

\bibitem{Park2008}
Park SH, et~al. (2008) The variation of relative magnetic helicity around major
  flares.
\newblock {\em The Astrophysical Journal} 686(2):1397.

\bibitem{Kontogiannis2017}
Kontogiannis I, Georgoulis MK, Park SH, Guerra JA (2017) Non-neutralized
  electric currents in solar active regions and flare productivity.
\newblock {\em Solar Physics} 292(11):159.

\bibitem{Sun2017}
Sun X, Hoeksema JT, Liu Y, Kazachenko M, Chen R (2017) Investigating the
  magnetic imprints of major solar eruptions with SDO/HMI high-cadence vector
  magnetograms.
\newblock {\em The Astrophysical Journal} 839(1):67.

\bibitem{Fisher2012}
Fisher GH, Bercik DJ, Welsch BT, Hudson HS (2012) Global forces in eruptive
  solar flares: The Lorentz force acting on the solar atmosphere and the solar
  interior.
\newblock {\em Solar Physics} 277(1):59--76.

\bibitem{Hoeksema2014}
Hoeksema JT, et~al. (2014) The Helioseismic and Magnetic Imager (HMI) vector magnetic field pipeline: Overview and performance.
\newblock {\em Solar Physics} 289(9):3483--3530.

\bibitem{Scherrer2012}
Scherrer PH, et~al. (2012) The Helioseismic and Magnetic Imager (HMI) investigation for the Solar Dynamics Observatory (SDO).
\newblock {\em Solar Physics} 275(1):207--227.

\bibitem{Pesnell-etall2012}
Pesnell WD, Thompson BJ, Chamberlin PC (2012) The Solar Dynamics Observatory  (SDO).
\newblock {\em Solar Physics} 275:3--15.

\bibitem{Ahmed2013}
Ahmed OW, et~al. (2013) Solar flare prediction using advanced feature
  extraction, machine learning, and feature selection.
\newblock {\em Solar Physics} 283(1):157--175.

\bibitem{bobraflareprediction}
Bobra MG, Couvidat S (2015) Solar flare prediction using SDO/HMI vector
  magnetic field data with a machine-learning algorithm.
\newblock {\em The Astrophysical Journal} 798(2):135.

\bibitem{Florios2018}
Florios K, et~al. (2018) Forecasting solar flares using magnetogram-based
  predictors and machine learning.
\newblock {\em Solar Physics} 293(2):28.

\bibitem{Jonas2018}
Jonas E, Bobra M, Shankar V, Todd~Hoeksema J, Recht B (2018) Flare prediction  using photospheric and coronal image data.
\newblock {\em Solar Physics} 293(3):48.

\bibitem{Raboonik2016}
Raboonik A, Safari H, Alipour N, Wheatland MS (2017) Prediction of solar flares using unique signatures of magnetic field images.
\newblock {\em The Astrophysical Journal} 834(1):11.

\bibitem{Nishizuka2017}
Nishizuka N, et~al. (2017) Solar flare prediction model with three
  machine-learning algorithms using ultraviolet brightening and vector
  magnetograms.
\newblock {\em The Astrophysical Journal} 835(2):156.

\bibitem{Huang2018}
Huang X, et~al. (2018) Deep learning based solar flare forecasting model. i.
  results for line-of-sight magnetograms.
\newblock {\em The Astrophysical Journal} 856(1):7.

\bibitem{Bobra2014}
Bobra MG, et~al. (2014) The helioseismic and magnetic imager (HMI) vector magnetic field pipeline: SHARPs -- space-weather HMI active region patches.
\newblock {\em Solar Physics} 289(9):3549--3578.

\bibitem{Wheatland2002}
Wheatland M, Litvinenko Y (2002) Understanding solar flare waiting-time
  distributions.
\newblock {\em Solar Physics} 211(1):255--274.

\bibitem{Longcope2000}
Longcope DW, Welsch BT (2000) A model for the emergence of a twisted magnetic
  flux tube.
\newblock {\em The Astrophysical Journal} 545(2):1089.

\bibitem{hamdi}
{Hamdi} SM, {Kempton} D, {Ma} R, {Boubrahimi} SF, {Angryk} RA (2017) A time
  series classification-based approach for solar flare prediction in {\em 2017
  IEEE International Conference on Big Data (Big Data)}.
\newblock pp. 2543--2551.

\end{thebibliography}
%\bibliographystyle{pnas-new}

\end{document}